# Relativistic Dynamical Friction in the Weak Scattering Limit

*D. Syer*

Canadian Institute for Theoretical Astrophysics, MacLennan Labs, 60 St. George Street, Toronto M5S 1A7, Ontario.

## Abstract

A test mass, $M$, moving through an ambient medium of light particles with lower average kinetic energy than itself suffers a deceleration caused by its scattering of the light particles. The phenomenon is usually referred to as dynamical friction. The velocity, $\mathbf{v}$, of the test mass decays on a timescale independent of $\mathbf{v}$ in the non-relativistic case. We derive expressions for dynamical friction in the case that the test mass and the light particles are relativistic, and that the scattering is weak (with impact parameter, $b \gg M$). In the case that the light particles are ultra-relativistic, and isotropic in the frame in which $M$ moves with velocity $v$, we find an explicit expression for the dynamical friction. The well known factor of 2 correcting the Newtonian scattering of photons to give the Einstein angle, $4M/b$, has the largest effect on the resulting friction, which is modified by a factor of roughly $16/3\gamma_v$ over the simple non-relativistic case. In the non-relativistic case, the largest contribution to the friction comes from light particles moving slower than $v$. We find that this is not the case for ultra-relativistic scattering, essentially because the scattering angle is independent of $\mathbf{v}$. Some astrophysical implications are discussed.

## 1 Introduction

Consider a test mass, $M$, moving through a medium of light particles. It will be useful to define the 'lab frame' as the one in which $M$ moves with velocity $\mathbf{v}$. In what follows we shall adopt natural units in which $G = c = 1$. Chandrasekhar (1943) derived a formula for the dynamical friction on $M$, which may be written

$$\frac{d\mathbf{v}}{dt} = -16\pi^2 \Lambda M m \frac{\int_0^v f(w) w^2\, dw}{v^3} \mathbf{v}, \qquad (1.1)$$

where $\Lambda$ is the Coulomb logarithm, and $f(w)$ is the distribution function of light particles, assumed to be isotropic. In the case that $v$ is much less than the average value of $w$, and $f$ is Maxwellian with dispersion $\bar{w}$,

$$\frac{d\mathbf{v}}{dt} \simeq -4\pi \Lambda \frac{M\rho}{\bar{w}^3}\mathbf{v}. \tag{1.2}$$

One important feature of (1.1) is that only the light particles with $w < v$ contribute to the friction, to order $\Lambda$. If all the light particles were massless, there would be none with $w < v$, so a naive application of (1.1) would predict no friction at all in this order. Of course, we would not be surprised to find contributions at other orders, but in the relativistic case $\Lambda$ can be very large, so it will be important to investigate the precise form of the dynamical friction. We shall show in the next section that the friction is actually enhanced if the light particles are ultra-relativistic. Our derivation is closer to that of Binney and Tremaine (1987) than to the original work by Chandrasekhar.

## 2 Dynamical Friction

Consider an event in which a light particle is scattered through a small angle, $\chi$, by gravitational interaction with $M$. The velocity change imparted to $M$ in a single collision is of order $m/M$ times that imparted to $m$. In the absence of a solution to the full relativistic two-body problem, we only know how to solve for the motion of $m$ in the limit that this quantity is small, $m/M \ll 1$. The motion of $M$ will be inferred by demanding conservation of momentum. Let us define the fictitious inertial frame, **I**, in which $M$ is stationary, and identify it with the frame in which $M$ is actually stationary at the beginning of the scattering event (i.e. when $m$ is at infinity). In this frame

$$\chi = \frac{2M}{b}\left(1 + \frac{1}{v_{\text{rel}}^2}\right) \tag{2.1}$$

(Misner, Thorne and Wheeler 1973) where $b$ is the impact parameter, and $v_{\text{rel}}$ is the relative velocity in **I**. This relative velocity is given by

$$v_{\text{rel}}^2 = \frac{|\mathbf{w}-\mathbf{v}|^2 - |\mathbf{v}\times\mathbf{w}|^2}{(1-\mathbf{v}.\mathbf{w})^2}. \tag{2.2}$$

In manifestly covariant form

$$v_{\text{rel}}^2 = \frac{(w_\mu v^\mu)^2 - 1}{(w_\mu v^\mu)^2}. \tag{2.3}$$

As $w \to 1$, we find that $v_{\text{rel}} \to 1$; and in the limit that both $w$ and $v$ are much less than the speed of light, $v_{\text{rel}} \to |\mathbf{w}-\mathbf{v}|$.

Now consider the scattering as viewed in the lab frame. We can write it as the product of a Lorentz boost, $\lambda$, into **I**, followed by a rotation in 3-space by an angle $\chi$, and finally the inverse boost back to the lab frame, $\lambda^{-1}$. The 4-momentum, $q^\mu$, of the light particle is changed by an amount $\Delta q^\mu$ given symbolically by

$$\Delta q^\mu = \left(\lambda^{-1} R(\chi)\lambda - 1\right) q^\mu. \tag{2.4}$$

We shall use a convenient notation in which only two space dimensions are retained—one parallel to **v**, and one perpendicular to it. (By symmetry, only the first of these contributes to the dynamical friction, but the other is retained for clarity.) Thus we write

$$q^\mu = \gamma_w m \begin{pmatrix} 1 \\ w\cos\theta \\ w\sin\theta \end{pmatrix}, \tag{2.5}$$

where $\theta$ is the angle between **v** and **w**. For additional brevity, we define $C = \cos\theta$ and $S = \sin\theta$. The boost $\lambda$ is thus written

$$\gamma_w m \begin{pmatrix} 1 \\ wC \\ wS \end{pmatrix} \to \gamma_w m \begin{pmatrix} \gamma_v(1 - vwC) \\ \gamma_v(wC - v) \\ wS \end{pmatrix}, \tag{2.6}$$

followed by the rotation

$$\to \gamma_w m \begin{pmatrix} \gamma_v(1 - vwC) \\ \gamma_v(wC - v)c - wSs \\ wSc + \gamma_v(wC - v)s \end{pmatrix}, \tag{2.7}$$

where $c = \cos\chi$ and $s = \sin\chi$. The final boost back to the lab frame takes $q^\mu$ to

$$\to \gamma_w m \begin{pmatrix} \gamma_v^2(1 - vwC + v(wC - v)c) - \gamma_v vwSs \\ \gamma_v^2((wC - v)c + v(1 - wvC)) - \gamma_v wSs \\ wSc + \gamma_v(wC - v)s \end{pmatrix}. \tag{2.8}$$

Thus we have, for $\chi$ small

$$\Delta q^\mu = \gamma_w m \begin{pmatrix} \gamma_v^2 v(wC - v)\chi^2/2 - \gamma_v vwS\chi \\ \gamma_v^2(wC - v)\chi^2/2 - \gamma_v wS\chi \\ wS\chi^2/2 + \gamma_v(wC - v)\chi \end{pmatrix}. \tag{2.9}$$

We now proceed to find the average rate of change of the momentum in the light particles, by averaging (2.9) over all possible impacts. If $M$ were stationary, the number of collisions in the lab frame in a time interval $dt$, in the impact range $[b, b + db]$, would given by

$$dN = nw\,2\pi b\,db\,dt, \tag{2.10}$$

where $n$ is the number density of light particles. To write $dN$ in a Lorentz invariant way, first we note that in terms of the proper time interval, $dt_0$, in **I**,

$$dt = \gamma_v dt_0. \tag{2.11}$$

Next, we examine the case that **v** is parallel to **w**, in which

$$dN = \gamma_v \gamma_w(w - v)n_0\,2\pi b\,db\,dt_0, \tag{2.12}$$

where $n_0$ is the proper number density of light particles in their rest frame. In terms of proper quantities and four vectors, and so in Lorentz invariant form, equation (2.12) may be written

$$dN = \left[(w_\mu v^\mu)^2 - 1\right]^{1/2} n_0 \, 2\pi b \, db \, dt_0. \tag{2.13}$$

This may in turn be written in terms of lab frame quantities as

$$dN = nV \, 2\pi b \, db \, dt, \tag{2.14}$$

where

$$V = \left[|\mathbf{w} - \mathbf{v}|^2 - |\mathbf{w} \times \mathbf{v}|^2\right]^{1/2}. \tag{2.15}$$

Suppose that $n$ is given by a distribution function $f(\mathbf{w})$, such that the density of light particles is

$$\rho = \int \gamma_w m f(\mathbf{w}) \, d^3\mathbf{w}, \tag{2.16}$$

then we find that

$$\frac{d\bar{q}^\mu}{dt} = \int 2\pi b \, db \, \Delta q^\mu \, V \, f(\mathbf{w}) d^3\mathbf{w}. \tag{2.17}$$

Integrating over $b$ cancels the contributions proportional to $\chi$ by symmetry (the tedious inclusion of the third space dimension would make this explicit). Thus

$$\frac{d\bar{q}^\mu}{dt} = \int 4\pi \Lambda M^2 \gamma_w m \begin{pmatrix} \gamma_v^2 v(wC - v) \\ \gamma_v^2(wC - v) \\ wS \end{pmatrix} \left(1 + \frac{1}{v_{\rm rel}^2}\right)^2 V f(\mathbf{w}) d^3\mathbf{w}, \tag{2.18}$$

where $\Lambda = \ln(b_{\max}/b_{\min})$. Since we are only considering weak scattering, $\Lambda$ is independent of $\mathbf{w}$. For a discussion of a natural choice for $b_{\min}$ in the case that $M$ is an extended distribution of mass see White (1976). We now concentrate on equation (2.18) in two interesting limits: first we re-derive the Chandrasekhar formula; and then we find the ultra-relativistic limit as $w \to 1$ for general $v$.

In the limit that $w$ and $v$ are much less than the speed of light, the 3-space part of (2.18) is equivalent to

$$\frac{d\bar{\mathbf{q}}}{dt} = \int 4\pi \Lambda M^2 m \frac{\mathbf{w} - \mathbf{v}}{|\mathbf{w} - \mathbf{v}|^3} f(\mathbf{w}) d^3\mathbf{w}.$$

For an isotropic distribution $f$,

$$\frac{d\bar{\mathbf{q}}}{dt} = \frac{4\pi \Lambda M^2 m \mathbf{v}}{v^3} \int_{w<v} f(\mathbf{w}) \, d^3\mathbf{w}, \tag{2.19}$$

which is the usual Chandrasekhar (1943) result.

In the limit that $w \to 1$, we must also demand that $m \to 0$. We write $f$ in terms of $\mathbf{q}$, by letting $\gamma_w m \to q$, whence

$$\frac{d\bar{q}^\mu}{dt} = \int 16\pi \Lambda M^2 \begin{pmatrix} \gamma_v^2 v(C - v) \\ \gamma_v^2(C - v) \\ S \end{pmatrix} (1 - vC) \, q f(\mathbf{q}) d^3\mathbf{q}.$$

For an isotropic distribution

$$\frac{d\bar{q}^\mu}{dt} = \frac{64}{3}\pi \Lambda M^2 \rho \gamma_v^2 v \begin{pmatrix} v \\ 1 \\ 0 \end{pmatrix}. \tag{2.20}$$

The momentum, $Q^\mu$, of $M$ will suffer an equal and opposite change, so

$$\frac{dQ^\mu}{dt} = -\frac{d\bar{q}^\mu}{dt}. \tag{2.21}$$

From this we may immediately find the interesting limit of the friction formula in which $M$ is a hard photon. Taking the limit $v \to 1$, $\gamma_v M \to Q$ in equation (2.20), we see that

$$\frac{d\bar{Q}^\mu}{dt} = \frac{64}{3}\pi \Lambda Q^2 \rho \begin{pmatrix} 1 \\ 1 \\ 0 \end{pmatrix}, \tag{2.22}$$

(which is very small in any astrophysical case).

Noting that

$$Q^\mu = \gamma_v M \begin{pmatrix} 1 \\ v \\ 0 \end{pmatrix}, \tag{2.23}$$

we derive

$$\frac{dQ^\mu}{dt} = \gamma_v^3 M \frac{dv}{dt} \begin{pmatrix} v \\ 1 \\ 0 \end{pmatrix}, \tag{2.24}$$

and thus that the energy and momentum parts of (2.18) are consistent, given that the component perpendicular to $\mathbf{v}$ must vanish by symmetry.

Thus finally, in the case $w \to 1$ and where $f$ is isotropic we have, from (2.20) and (2.24),

$$\frac{d\mathbf{v}}{dt} = -\frac{64}{3}\pi \Lambda \frac{M\rho}{\gamma_v}\mathbf{v}. \tag{2.25}$$

Re-inserting the dimensional constants $G$ and $c$ this is

$$\frac{d\mathbf{v}}{dt} = -\frac{64}{3}\pi \Lambda \frac{G^2 M\rho}{\gamma_v c^3}\mathbf{v}. \tag{2.26}$$

Comparing equations (2.25) and (1.2) we see that, (i), the fact that all the light particles have higher velocities than $M$ does not reduce the dynamical friction; and (ii), the rate of change of $v$ due to dynamical friction is modified by a factor of roughly $16/3\gamma_v$ over the simple non-relativistic case.

## 3 Discussion

We should be aware of the effect of direct collisions between $M$ and the light particles. If the light particles are photons, this corresponds to the radiation pressure force on $M$, assuming it is optically thick. Suppose that $M$ was a 'shiny disc' of radius $a$, meaning that, in a direct collision, the light particles are elastically reflected in the direction of $v$. Going through a similar derivation to the above we would find that, in the limit $w \to 1$,

$$\frac{d\mathbf{v}}{dt} = -\frac{16}{3}\pi \frac{\rho}{\gamma_v M}\mathbf{v}. \tag{3.1}$$

The ratio of the dynamical friction rate to the direct friction rate is $4\Lambda M^2/a^2 \sim \Lambda v_{\text{esc}}^4$, where $v_{\text{esc}} = (2M/a)^{1/2}$ is the escape speed from the surface of $M$.

For objects, such as stars, $v_{\text{esc}} \ll 1$ and hence dynamical friction by photons can never be important. There are, however, at least two cases in which dynamical friction against relativistic light particles could be interesting: if $v_{\text{esc}} \sim 1$; or if the optical depth, $\tau$, of $M$ to the light particles satisfied $\tau \ll v_{\text{esc}}^2$.

If $v_{\text{esc}} \sim 1$ then the weak-scattering approximation would no longer be valid, but actually the dynamical friction would be greater (the effective $\Lambda$ would be vary large). Thus neutron stars and black holes would feel a greater force due to dynamical friction than due to radiation pressure, if immersed in a background of photons of lower energy. The ratio of the radiation pressure to the gravitational force on a black hole of mass $M$ in the vicinity of an object radiating at its Eddington limit is

$$\frac{F_{\text{radiation}}}{F_{\text{gravity}}} \sim L_M \sim 10^{-15}\frac{M}{M_\odot}, \tag{3.2}$$

where $L_M$ is the Eddington luminosity of $M$ itself in natural units. Thus for radiation pressure to have an appreciable effect on the orbit of $M$ in a Hubble time, we require that

$$10^{-15}\frac{M}{M_\odot}\frac{t_{\text{Hubble}}}{t_{\text{orbit}}} \gtrsim 1, \tag{3.3}$$

where $t_{\text{orbit}}$ is the period of the orbit. Writing $t_{\text{orbit}}$ in terms of the separation $\eta M$ of two equal mass black holes, the inequality (3.3) becomes

$$\eta^{3/2} \lesssim 10^7, \tag{3.4}$$

independent of $M$. For $\eta \sim 1$ gravitational radiation would take over as the dominant effect on the orbit, but gravitational radiation decreases quickly with increasing separation, so radiation pressure might be important on timescales as short as $10^6$ years. Dynamical friction would be comparable or larger.

If the light particles were neutrinos, then a star would be very optically thin and direct collisions would not provide any friction at all. Neutrino balls (Holdom 1987, Holdom and Malaney 1994) are an example in which the neutrino density would be high enough to affect the orbit of a star. In such objects the neutrinos are not strictly massless, but they are degenerate, and so they are virtually all ultra-relativistic.